\documentclass[conference]{IEEEtran}
\IEEEoverridecommandlockouts
\usepackage{cite}
\usepackage{amsmath,amssymb,amsfonts}
\usepackage{algorithmic}
\usepackage{graphicx}
\usepackage{textcomp}
\usepackage{xcolor}
\def\BibTeX{{\rm B\kern-.05em{\sc i\kern-.025em b}\kern-.08em
    T\kern-.1667em\lower.7ex\hbox{E}\kern-.125emX}}
\begin{document}

\title{Character Simulation Using Imitation Learning With Game Engine Physics\\

}

\author{\IEEEauthorblockN{João Rodrigues}
\IEEEauthorblockA{\textit{NOVA LINCS} \\
\textit{NOVA School of Science and Technology}\\
Caparica, Portugal \\
jfpe.rodrigues@campus.fct.unl.pt}
\and
\IEEEauthorblockN{Rui Nóbrega}
\IEEEauthorblockA{\textit{NOVA LINCS} \\
\textit{NOVA School of Science and Technology}\\
Caparica, Portugal \\
rui.nobrega@fct.unl.pt}
}

\maketitle

\begin{abstract}
Creating visual 3D sensing characters that interact with AI peers and virtual environments can be a difficult task for those with less experience in using learning algorithms or creating visual environments to execute an agent-based simulation. In this paper, the use of game engines as a tool to create and execute graphic simulations with 3D sensing characters is being explored with plugins such as ML-Agents for the Unity3D game engine. This allows the simulation of agents using off-the-shelf algorithms and using the game engine's motor for the visualizations of these agents. We explore the use of these tools to create visual bots for games, and teach them how to play the game until they reach a level where they can serve as adversaries for real-life players in interactive games.
\end{abstract}


\section{Introduction}
Games have always used interactive characters and Non-Playable Characters (NPCs) to make the game more interesting, appealing and relatable. Agent and Character Simulation has been used over time for many different purposes\cite{Macal2009}, going from scientific research, such as predicting the spread of pandemics\cite{Khalil2012}, to leisure\cite{Juliani2018} with games having NPCs to add more interactivity or make the game harder. Using these simulations requires a vast knowledge of the different AI, machine learning, and deep learning algorithms needed for these simulations. 
Additionally, many of these simulations also need graphic environments to show the results with rendering techniques and physics simulations. All these techniques require specialized knowledge that for a computer graphics developer may not be available, specially in the AI and Machine Learning fields.

Another factor that is also present in agent simulation, is the need for create sensing characters. Many of the Agent-Based Modelling (ABM) and Simulation scenarios require characters that interact with the environments
or with their peers. To achieve this, the creation of characters that can sense
their surroundings by simulating vision, touch, and distance among other factors is needed. Creating these senses from scratch also needs advanced programming knowledge  such as ray tracing for vision and distance, or physics for touch.

All these needs constitute a problem for those who want to create character simulations but do not have the technical expertise to develop all the aforementioned technologies. 
To solve the problems created by the need for a graphical environment with sensing characters, it can be interesting to use game engines to visualize the environment and simulate the senses. Being in constant development to facilitate the process of creating games, game engines seem to be a good tool to create and train characters and environments since they already have render and physics engines that implement virtual environment rules sush as ray cast distance, gravity or colisions.

Besides the use of game engines, over the years many tools have been created to facilitate the creation and execution of these simulations by providing utilities that facilitate the testing and sharing of different machine and deep learning algorithms such as OpenAI Gym~\cite{Brockman2016} and PettingZoo~\cite{Terry2020}. One tool that has special interest when considering the use of game engines is the ML-Agents plugin~\cite{Juliani2018}, since it was created specifically to integrate with the Unity3D game engine and aims to ease the creation and learning of game bots.

This last tool, the ML-Agents, contains a utility that can be helpful in teaching bots how to play which is the demonstration recorder. This tool allows recording the steps of a human playing in the same environment as the bots, collect the observations the bot would receive during the playtime, and generate a demo file that can be later used to teach bots how to play.

In this paper we propose to create a set of game scenarios and use them to teach bots how to play. More than this, we will study the use of human play recordings as a way of teaching the bots how to play. Afterwards we aim to compare if this method shows results similar to human level of play.

To have a better grasp of how game engines can be used for a better graphical interface for agent-based simulation we are going to look at some of the related work done in these areas to understand better a few of the concepts of agent simulation, game engines, and some libraries that can be used as a tool to connect both. After this, the idea and concepts behind the work done are going to be exposed for a better understanding of how ML-Agents works, and how we can use it to improve learning bots' performance. 




\section{Related Work}




A game engine’s main use is to produce games by rendering its graphics, producing music and sound effects, and allowing external manipulation of the scene by reading inputs from an input device. Usually, a game engine is made up of Rendering Engine, Animation Engine, Physics Engine, Artificial Intelligence Engine, Network Engine, 3D Sound Engine, and a Map Editor~\cite{Noh2016}.

Although the main purpose of a game engine is, as its name says, to create games, its features allow for more uses such as animations and simulations. The render, animation, and physics engines can facilitate the production of animation. With these engines one can just pick up a scene to animate, code the intended movements, and then just play out the result which will follow physics without having to animate frame by frame every movement, and collision~\cite{Noh2016}. In the same line as animation, these tools can be used to produce simulations. 

What makes the use of a game engine alluring for animations and simulations
is how it allows a user to view these creations, without having to make low-level code to render and animate the scene since the game engine deals with this internally\cite{Bak2020}. For agent simulation, it interests the most how a game engine allows manipulating the motion of the agent’s models and how to create a simulation environment for the agents to play out simply by adding game objects to a scene, and manipulating its properties with no need for extensive lines of code describing the shape, physics, and motion of every object in the
simulation. Some of the works done in using game engines for simulations are~\cite{Bartneck2015,Pasternak2018,Carlos2008}.

\subsection{Graphical Agent Simulation}

Graphical Agent-Based Modeling and simulation is a paradigm in which simulated humans, animals, or other forms of beings are modeled as agents that interact with their peers as well as their environment~\cite{Klugl2012}. In these agent-based simulations, sometimes called multi-agent simulation, the environment plays a crucial role since it influences all the agents and their interactions and therefore must be carefully taken into account.

In order to create an agent-based simulation, four elements have to be taken into consideration: the set of agents, set of interactions, the environment and the simulation infrastructure~\cite{Klugl2004}.

Self-play reinforcement learning is when agents learn by exploring and playing with
themselves. It has seen success when applied in many game scenarios, however, the
process for self-play learning is unstable and more sample-inefficient than general reinforcement learning, especially when used in imperfect information games~\cite{Baker2019}. Multi-agents auto-curricula 
have been used to solve the various type of multiplayer games both in classic discrete games,  such as Backgammon~\cite{10.1145/203330} and  Chess~\cite{DBLP:journals/corr/abs-1712-01815}, and continuous real-time games like Dota and Starcraft. Illustrative examples of works done with agents simulation and self-play reinforcement learning using visual scenarios are given by~\cite{Baker2019,Yang2020,OpenAI_dota, Vinyals2019}.

\subsection{Agent Simulation Libraries}

In order to create sensing agents for game engine environments, it is crucial to find a library that allows simulating the agent’s actions without the need to write complex algorithms. Luckily, nowadays, there is an increasing number of these that are open source and allow for indiscriminate integration in any kind of simulation as long as it complies with the library specifications.


Unity Machine Learning Agents~\cite{Juliani2018} (ML-Agents) is a toolkit that allows games and simulation to serve as environments for training intelligent agents. This library is specific for Unity’s 3D game engine and provides implementations of state-of-the-art that enable game developers and researchers to easily train agents for 2D, 3D, and VR/AR games. With ML-Agents, agents can be trained using reinforcement learning, imitation learning, neuroevolution, and some other methods with the provided simple-to-use Python API.

\section{Simulation and Training Workflow}

As mentioned before, the goal is to get records of humans playing and using the obtained data to train bots. 

In the context of this paper these records will be called demonstrations. A demonstration is a data structure that saves the observations humans obtained at each moment, and maps it to the actions they took, so that it is possible for bots to decide their actions, by finding the actions in the demonstrations that have similar observations to what they have at that point.

To better understand how the demonstration records can be used to improve the learning speed of bots, first it is good to understand how to create scenarios with learning bots and how ML-Agents teach them. Creating the scenarios is quite simple for those with game development experience. The first step is to envision the desired interactive scene, then, build that scene in Unity3D by placing different game objects and scripts that control the desired interactions.

\begin{figure}[b]
    \centering
    \includegraphics[width=1\linewidth]{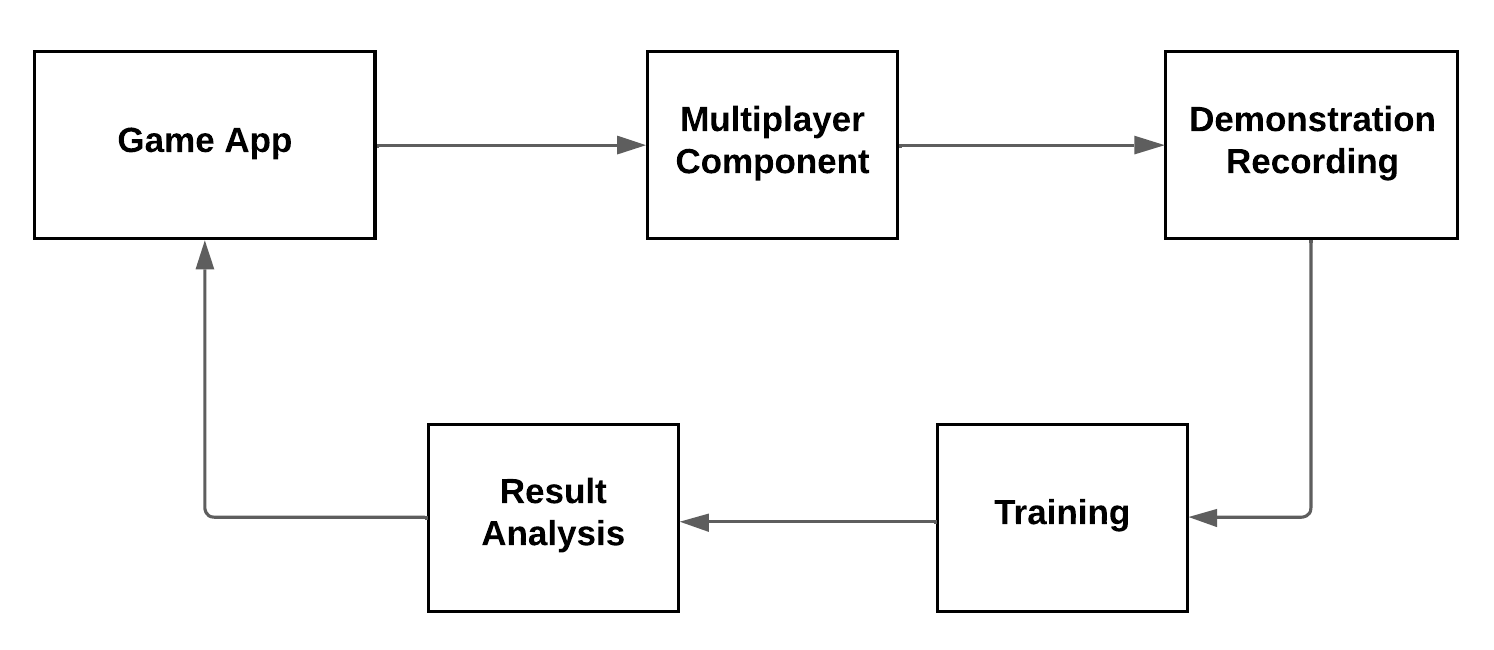}
    \caption{Demonstration Record process flowchart}
    \label{fig:flowchart}
\end{figure}

After this, each bot connects with the ML-Agents interface, its behaviour takes into consideration the actions that it receives from the system and the observations from the environment it interacts with. Each bot sends a set of variables and values that constitute an observation, and a hit history from ray-traces from the bot to the world.


Outside of Unity3D, the ML-Agents library is executed using a configuration file that defines: run id, flags and the specific training algorithm. The training algorithms can be split into two categories, the Reinforcement Learning algorithms, and the Imitation Algorithms. For RL algorithms, ML-Agents offers Proximal Policy Optimization (PPO) and Soft Actor-Critic (SAC) for single agent scenarios, Multi-Agent Posthumous Credit Assignment (MA-POCA) for multi agent cooperative/competitive scenarios, and self-play which can be used in both situations.

For the imitation algorithms, the ML-Agents library\footnote{ML-Agents, https://github.com/Unity-Technologies/ML-Agents/blob/main/docs/Training-ML-Agents.md} offers Behavioral Cloning and Generative Adversarial Imitation Learning (GAIL) which use the demonstration records to perform imitation learning. These algorithms can be used together as a better way to reinforce the imitation learning, and can also be used with RL Algorithms as a way to learn past the demonstrations and use them only as a base for learning.

\begin{figure}[b]
    \centering
    \includegraphics[width=1\linewidth]{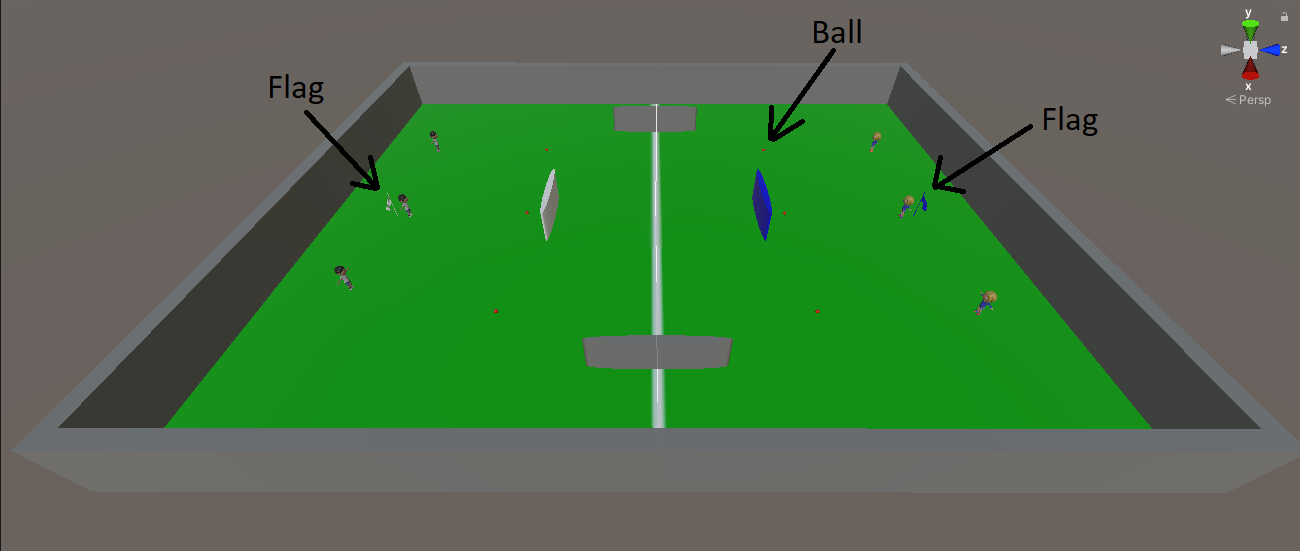}
    \caption{Capture the Flag Scene where players have to capture enemy flag and retrieve it to their own field}
    \label{fig:CTF_Scene}
\end{figure}

Now that a broad look at how ML-Agents can be used to create self-learning bots in Unity3D was seen, we can focus on the question of how can demonstration records help bots learn. First, let's look 
at what exactly are these demonstration records. With the ML-Agents package in unity, it is possible to attach a demonstration recorder to an agent. When attached, this recorder will save a data structure with all the actions the humans who are playing did mapped to all the variables and ray trace hits provided as observations at each action point. This forms a demonstration for imitation learning because the agents can then use these data structures to decide what actions to take, by comparing their observations with the observations present on the demo file, and see which one looks the most similar and perform the action took with that observation on the demo. This means that to generate demos that can be used for learning, we can have human users playing the scenarios and complete the tasks to serve as inspiration for the bots. This is possible because the ML-Agents interface allows an heuristic function which can be used to create human controls for the bots.

The idea with this is that for more complex scenarios that take very long for bots to learn, and that the tuning of the configuration file becomes harder, we can take advantage of these demonstrations to accelerate the process and create usable bots faster. For this, first we have to create a game that is designed for both the agents and users to play it out. We also need to define the observations and rewards the agent gets. If this game involves multiple agents simultaneously it has to support multiplayer so multiple users can also play it out simultaneously. After all this is done, we proceed to have users play the game to record the demonstrations. The more users and the more time the better. Upon collecting all the demonstrations, the training of the bots can be started. Then the final result of the bots can be analyzed, and if it's still not up to expectations, the process can be repeated to collect more demonstrations with the possibility of using the current bots as adversaries to the users. A flowchart of this process can be seen in Figure \ref{fig:flowchart}.

To test this idea of using demonstration records to improve the learning speed of bots we decided the use a game of capture the flag in a 3\textit{vs}3 player competitive environment as seen in Figure \ref{fig:CTF_Scene}. This game was chosen because we wanted a simultaneously competitive/cooperative game, with a decent amount of complexity, so that the performance level differences are easily distinguishable, but also it's not extremely hard for bots and humans to learn how to play the game.

\begin{figure}[b]
    \centering
    \includegraphics[width=1\linewidth]{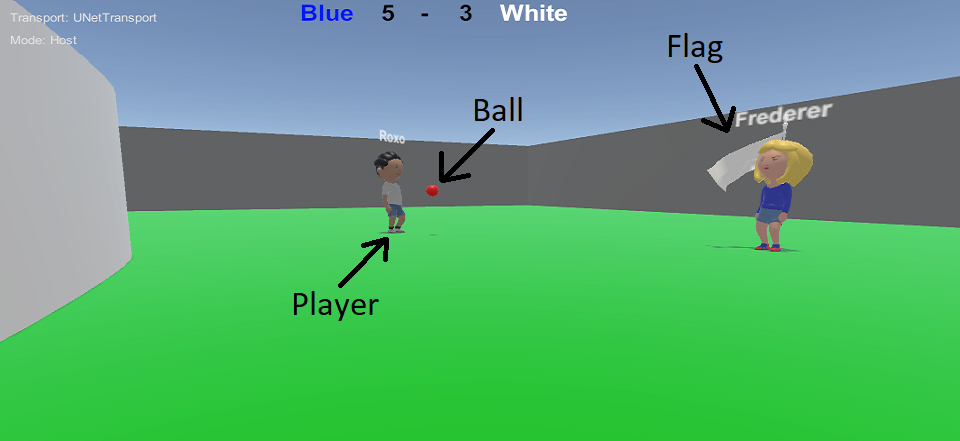}
    \caption{Game-play screenshot displaying all Network Objects}
    \label{fig:gameplay}
\end{figure}

In this game, 2 teams (team Blue and team White) compete to try and capture the enemy flag, and take it all the way back to their teams field location. Each iteration of capturing a flag and retrieving it to the teams field is, for training purposes, considered a round. In addition to this, the field is spread with a few walls to create obstacles for the players, and balls the players can catch and throw to attempt to hit other players. If a players gets with while holding a flag, they will drop it and be stunned for 3 seconds. While a flag is dropped out of place, both teams can attempt to catch it, to either recapture it or make it go back to spawn depending if it is the enemy or ally flag respectively. An image illustrating the described game can be seen in Figure \ref{fig:CTF_Scene}. This game supports that multiple users play at the same time,
meaning that it is necessary to implement a multiplayer component for this game. This is needed so that the human users can play and record demonstrations at a competitive level by playing other humans.

\section{Preliminary Results}

For this initial experiment, 3 play sessions with human users were played to record demonstrations. During these play sessions a total of 10 demonstrations were obtained. After the recording sessions the bots were trained several times using different numbers of demos. 

In these initial results, it was possible to see that the bots got better as more demonstrations were used. The amount of time needed to complete each round of the game dropped from an average of 517 seconds with bots trained using 1 demo to 165 seconds when 10 demos were used. The number of draws (when neither team is able to win after 1000 seconds) also lowers as more demos are used, having a ratio of 20\% of games drawn with 1 demo and only 7\% of games drawn with 10 demos. 

During the recording sessions, it was noted that the team playing the left side of the field (Blue team) won overall 68.9\% of all rounds played, having more rounds won in all three of the play sessions. This dominance was also translated to the bots and had increased strength as more demonstrations were used. When only 1 demo was used to train the bots, the blue team won 57\% of games having a slight superiority, but when the 10 demonstrations were used to train the bots, the blue team won 81\% having a huge superiority, even bigger than the one the human blue team had.

Looking back at times per round, we analyzed how quickly each of the teams finished the game in their winning rounds. Against expectations, on their wins the White team had an average of 50 seconds and were able to finish the game, 50 seconds faster than the Blue team which had an average time on wins of 100 seconds. A first explanation for this event is that the White team is only able to win on ideal circumstances, while the blue team can find ways to win even if the game drags a bit longer.

\section{Conclusions}

In this paper, we present a method to create game engine self-learning bots in Unity3D that use imitation algorithms. This technique was achieved using the described ML-Agents plugin. Then we focused on the hypothesis that imitation learning methods using demonstration records can be a good way to create bots for games with complex environments. For this, we created a 3\textit{vs}3 game of capture the flag and had multiple human users record demonstrations simultaneously.

With the results obtained we conclude that imitation learning can be a good way to train game-ready bots and that further investigation is necessary. Firstly, the recording of many more demonstration recordings is needed so that it is possible to analyze the level of the bots when a big sample of demonstrations is available and compare it to the human level by games between humans and bots. Following the increase in the sample size, it will also be interesting to see the results if a more balanced set of results is used and see if then both teams have a similar win ratio or if one team keeps dominating regardless. Lastly, it is left to expand this method of teaching to other games and see the results and the level of trained bots in games with varying levels of complexity and see what variables tend to facilitate or difficult the bots learning process through imitation.

\section*{Acknowledgment}
This work was supported by NOVA.ID.FCT/NOVA LINCS (UIDB/04516/2020)

\bibliographystyle{IEEEtran}
\bibliography{conference_101719}

\vspace{12pt}

\end{document}